\documentclass[twocolumn,superscriptaddress,amsfont,amssymb,amsmath, showpacs,balancelastpage, nofootinbib]{revtex4-1}
\usepackage{amssymb,amsmath,mathtools,mathrsfs} 
\usepackage[normalem]{ulem}
\usepackage{enumerate,todonotes} 
\usepackage{placeins}
\usepackage{hyperref}
\hypersetup{
    colorlinks,
    linkcolor={red!75!black},
    citecolor={blue!75!black},
    urlcolor={blue!75!black}
}
\usepackage{float,todonotes}
\usepackage{setspace}
\usepackage{graphicx}
\usepackage{todonotes}
\usepackage{aas_macros}
\usepackage{xfrac}
\usepackage{tabularx}
\usepackage[caption=false]{subfig}
\usepackage{multirow}
\usepackage[utf8]{inputenc}
\usepackage{tikz-feynman}
\tikzfeynmanset{
  extra large/.style={
    /tikz/node distance=4cm,
    /graph drawing/node distance=5cm,
    /graph drawing/level distance=4cm,
    /graph drawing/sibling distance=6cm,
    /tikz/graphs/edges={very thick},
    /tikzfeynman/every dot@@/.append style={/tikz/minimum size=4mm},
    /tikzfeynman/every crossed dot@@/.append style={/tikz/minimum size=8mm},
    /tikzfeynman/every blob@@/.append style={/tikz/minimum size=2cm},
    /tikzfeynman/arrow size=3.2pt,
    /tikzfeynman/insertion/size=8pt,
  },
}

\newcommand{\be}{\begin{equation}}
\newcommand{\ee}{\end{equation}}

\newcommand{\MPl}{M_{\rm Pl}}
\newcommand{\G}{  \bar{G} }
\newcommand{\aB}{\alpha_B}
\newcommand{\aM}{\alpha_M}
\newcommand{\aT}{\alpha_T}
\newcommand{\aK}{\alpha_K}
\newcommand{\aI}{\alpha_i}

\newcommand{\comment}[1]{}
\newcommand{\nn}{\nonumber}

\newcolumntype{C}[1]{>{\centering\let\newline\\\arraybackslash\hspace{0pt}}m{#1}}

\begin{document}
\title{Positivity in the sky}

 \author{S.~Melville}
  \affiliation{DAMTP, University of Cambridge, Wilberforce Road, Cambridge CB3 0WA, U.K.}
  \affiliation{Emmanuel College, University of Cambridge, St Andrew’s Street, Cambridge CB2 3AP, U.K.}
  \author{J.~Noller}
  \affiliation{Institute for Theoretical Studies, ETH Z\"urich, Clausiusstrasse 47, 8092 Z\"urich, Switzerland}
  \affiliation{Institute for Particle Physics and Astrophysics, ETH Z\"urich, 8093 Z\"urich, Switzerland}

\begin{abstract}
\noindent Positivity bounds -- the consequences of requiring a unitary, causal, local UV completion -- place strong restrictions on theories of dark energy and/or modified gravity. 
We derive and investigate such bounds for Horndeski scalar-tensor theories and for the first time pair these bounds with a cosmological parameter estimation analysis, using CMB, redshift space distortion, matter power spectrum and BAO measurements from the Planck, SDSS/BOSS and 6dF surveys. Using positivity bounds as theoretical priors, we show that their inclusion in the parameter estimation significantly improves the constraints on dark energy/modified gravity parameters. 
Considering as an example a specific class of models, which are particularly well-suited to illustrate the constraining power of positivity bounds, we find that these bounds eliminate over $60\%$ of the previously allowed parameter space. 
We also discuss how combining positivity requirements with additional theoretical priors has the potential to further tighten these constraints: for instance also requiring a subluminal speed of gravitational waves eliminates all but $\lesssim 1\%$ of the previously allowed parameter space.
\end{abstract}

\date{\today}
\maketitle

\noindent Recently, significant progress has been made in developing parameterised approaches that allow model-independent precision-testing of our current leading theory of gravity, General Relativity (GR), as well as dark energy/modified gravity-related deviations away from it, in a (linear) cosmological setting \cite{Gubitosi:2012hu,Bloomfield:2012ff,Gleyzes:2014rba,Bellini:2014fua,Gleyzes:2013ooa,Kase:2014cwa,DeFelice:2015isa,Langlois:2017mxy,Lagos:2016wyv,Lagos:2017hdr}. 
Simultaneously, there have been advances in understanding what theoretical consistency criteria are required of {\it low-energy} Effective Field Theories (EFTs) to allow for a well-defined {\it high-energy} (UV) completion -- and what these so-called ``positivity bounds'' imply for (low-energy) theories of dark energy and modified gravity \cite{Adams:2006sv,Jenkins:2006ia,  Adams:2008hp, Nicolis:2009qm, Bellazzini:2014waa, Bellazzini:2015cra, Baumann:2015nta, Bellazzini:2016xrt, Cheung:2016yqr,Bonifacio:2016wcb,deRham:2017avq,deRham:2017imi,deRham:2017zjm,Bellazzini:2017fep,deRham:2017xox,deRham:2018qqo,Bellazzini:2019xts}.
While cosmological parameter constraints on deviations from GR have been computed using general parameterised approaches and a variety of (current and forecast) experimental data \cite{mcmc,BelliniParam,Hu:2013twa,Raveri:2014cka,Gleyzes:2015rua,Kreisch:2017uet,Zumalacarregui:2016pph,Alonso:2016suf,Arai:2017hxj,Frusciante:2018jzw,Reischke:2018ooh,Mancini:2018qtb,radstab}, 
positivity bounds have so far not been paired with any such observational constraints on gravity. Here we will do so for the first time and show that a holistic joint analysis, which takes into account both theoretical priors required by positivity and observational constraints from recent data, can significantly improve cosmological parameter constraints on deviations from GR.

Scalar-tensor (ST) theories -- minimal deviations from GR in the sense that they only introduce a single additional degree of freedom -- are at the heart of the parameterised approaches for dark energy and modified gravity that have been developed so far. Accordingly, we will consider Horndeski gravity \cite{Horndeski:1974wa,Deffayet:2011gz}, the most general Lorentz-invariant ST action that gives rise to second order equations of motion for the metric, $g_{\mu \nu}$, and for the additional scalar field, $\phi$. Specifically, this amounts to any linear superposition of the following four terms
\begin{align}
\mathcal{L}_2&=  \Lambda_2^4 ~G_2  ~, \quad\quad\quad\quad \mathcal{L}_3=  \Lambda_2^4 ~G_3 ~  [\Phi] ~,\nn \\
\mathcal{L}_4&=  \MPl^2 G_{4}  R +  \Lambda_2^4 ~ G_{4,X}  ~  \left( [\Phi]^2-[\Phi^2] \right) ~,   \label{lags}
 \\
 \mathcal{L}_5&=   \MPl^2 G_{5} G_{\mu\nu}\Phi^{\mu\nu} -\tfrac{1}{6}\Lambda_2^4  G_{5,X}  ([\Phi]^3 -3[\Phi][\Phi^2]+2[\Phi^3])  , \nn
 \end{align}
where second derivatives of $\phi$ enter via the dimensionless matrix $\Phi^{\mu}_{\;\; \nu} \equiv  \nabla^\mu \nabla_\nu\phi / \Lambda_3^3$, square brackets denote the trace, e.g. 
$ [\Phi^2] \equiv  \nabla^\mu\nabla_\nu\phi\nabla^\nu\nabla_\mu\phi / \Lambda_3^6 $, and the $G_i$ are free functions of $\phi$ and $\nabla^\mu\phi\nabla_\mu\phi$. 
Specifically, we have chosen to write the $G_i$ as functions of the dimensionless combinations $\phi/\Lambda_1$ and
$X \equiv - \tfrac{1}{2}  \nabla^\mu \phi \nabla_\mu \phi / \Lambda_2^4$, where the subscripts ``$,\phi$'' and ``$,X$'' denote derivatives with respect to these (dimensionless) arguments
and the constant mass scales $\Lambda_i$ are taken\footnote{
See Refs. \cite{Pirtskhalava:2015nla, radstab} for further discussion of this choice.
} to be $\Lambda_1 = \MPl$, $\Lambda_2^2 = \MPl H_0$, and $\Lambda_3^3 = \MPl H_0^2$. Here $\MPl$ is the (reduced) Planck mass and $H_0$ is the Hubble parameter today. 
From an EFT point of view, these represent the scales at which different sectors of the theory become \emph{strongly coupled}, defining a regime of validity beyond which trustworthy predictions can no longer be made.\footnote{Note that the near simultaneous detections of GW170817 and GRB 170817A \cite{PhysRevLett.119.161101,2041-8205-848-2-L14,2041-8205-848-2-L15,2041-8205-848-2-L13,2041-8205-848-2-L12} have also been used to significantly reduce the functional freedom in Horndeski gravity \cite{Baker:2017hug,Creminelli:2017sry,Sakstein:2017xjx,Ezquiaga:2017ekz}, in particular placing tight restrictions on $G_4 (X)$.
However, the frequencies of the merger are close to $\Lambda_3$, so additional assumptions about the UV physics are necessary to apply these bounds \cite{deRham:2018red} (also see Refs. \cite{Creminelli:2017sry,Creminelli:2018xsv} for related discussions). Our goal here is to remain as agnostic as possible about the UV physics, so we will not fix the speed of cosmological gravitational waves here.   
}
The full Horndeski theory can then be written as
\begin{align}\label{Horn_action}
S_{\rm H}=\int d^4x \sqrt{-g}\left\{\sum_{i=2}^5 {\cal L}_i [\phi,g_{\mu\nu}]\right\} \, .
\end{align}

In order to best illustrate the impact positivity bounds can have on cosmological parameter estimation, we will focus on a concrete example in the main text (and discuss the general case in the Appendix). Specifically, we consider the shift-symmetric part of the $\mathcal{L}_2$ and $\mathcal{L}_4$ pieces in \eqref{lags}, i.e.
\begin{align}\label{quarticA}
S &= \int d^4x \sqrt{-g} \Big\{ \Lambda_2^4 G_2(X) + \MPl^2 G_{4}(X) R \nn \\ 
&\qquad\qquad\qquad+ \Lambda_2^4 G_{4,X}(X) \left( [\Phi]^2-[\Phi^2] \right)\Big\},
\end{align}
and also allow for a small mass term, $- \tfrac{1}{2} m^2 \phi^2$.
We will see that this subclass of Horndeski theories is an excellent example of how current positivity bounds and observational constraints complement one another, but ultimately stress that this is a first step towards a more complete, integrated analysis: As more observational data become available and additional positivity bounds are computed in the future, we fully expect a much wider set of theories to be constrained increasingly tightly.
\\

\noindent \textit{Positivity bounds:} 
Since many of the terms in \eqref{lags} and \eqref{quarticA} are non-renormalizable, these theories must break down at high energies (typically around $\Lambda_3$). They are intended as an effective low energy description of some (potentially very complicated) underlying microphysics. 
Rather than trying to guess at this fundamental underlying theory, we will assume only that it is consistent with a ``standard'' Wilsonian field theory description -- one in which Lorentz invariance, unitarity (well-defined probabilities), analyticity (causality) and polynomial boundedness (locality) are respected. From these basic principles, one can construct a variety of constraints which the low energy parameters (here encoded in the $G_i$) must satisfy, known as ``positivity bounds''  \cite{Adams:2006sv,Nicolis:2009qm,Bellazzini:2016xrt,deRham:2017zjm}. The simplest of these concerns the tree-level scattering amplitude, $\mathcal{A}$, between two massive particles on a flat (Minkowski) background (see the Appendix for subtleties related to massless particles and non-trivial backgrounds). 
When expanded in powers of the center of mass energy, $s$, and the momentum transfer, $t$, 
\begin{equation}
\mathcal{A} (s,t) =  c_{ss} ~ \frac{s^2}{\Lambda_2^4}  + c_{sst} ~ \frac{ s^2 t}{\Lambda_3^6} +  ... \; , 
\label{Ast}
\end{equation}
the expansion coefficients must obey the bounds \cite{Adams:2006sv, Nicolis:2009qm, deRham:2017avq},
\begin{equation}
c_{ss} \geq 0 \, , \qquad c_{sst}  \geq  -  c_{ss} \; 3 \Lambda_3^4 / 2 \Lambda_2^4 \, , 
\label{cABposbounds}
\end{equation}
up to additional contributions suppressed by $\mathcal{O} ( m^2 / \Lambda_3^2)$.
Notionally, this corresponds to diagnosing whether it is possible (even in principle) for some new physics to enter at the scales $\Lambda_3$ and $\Lambda_2$ to restore unitarity in the full UV amplitude. 
If these bounds were violated, it would indicate that this new high energy physics is quite unlike any quantum field theory we know today\footnote{
For alternatives to the kind of ``standard'' Wilsonian UV completion considered here, see Refs.  \cite{Dvali:2010jz,Dvali:2010ns,Dvali:2011nj,Dvali:2011th,Vikman:2012bx,Kovner:2012yi,Keltner:2015xda}. 
}.

\begin{figure}[t]
\begin{center}
\includegraphics[trim={0.55cm 0.6cm 1.0cm 1.0cm},clip,width=.99\linewidth]{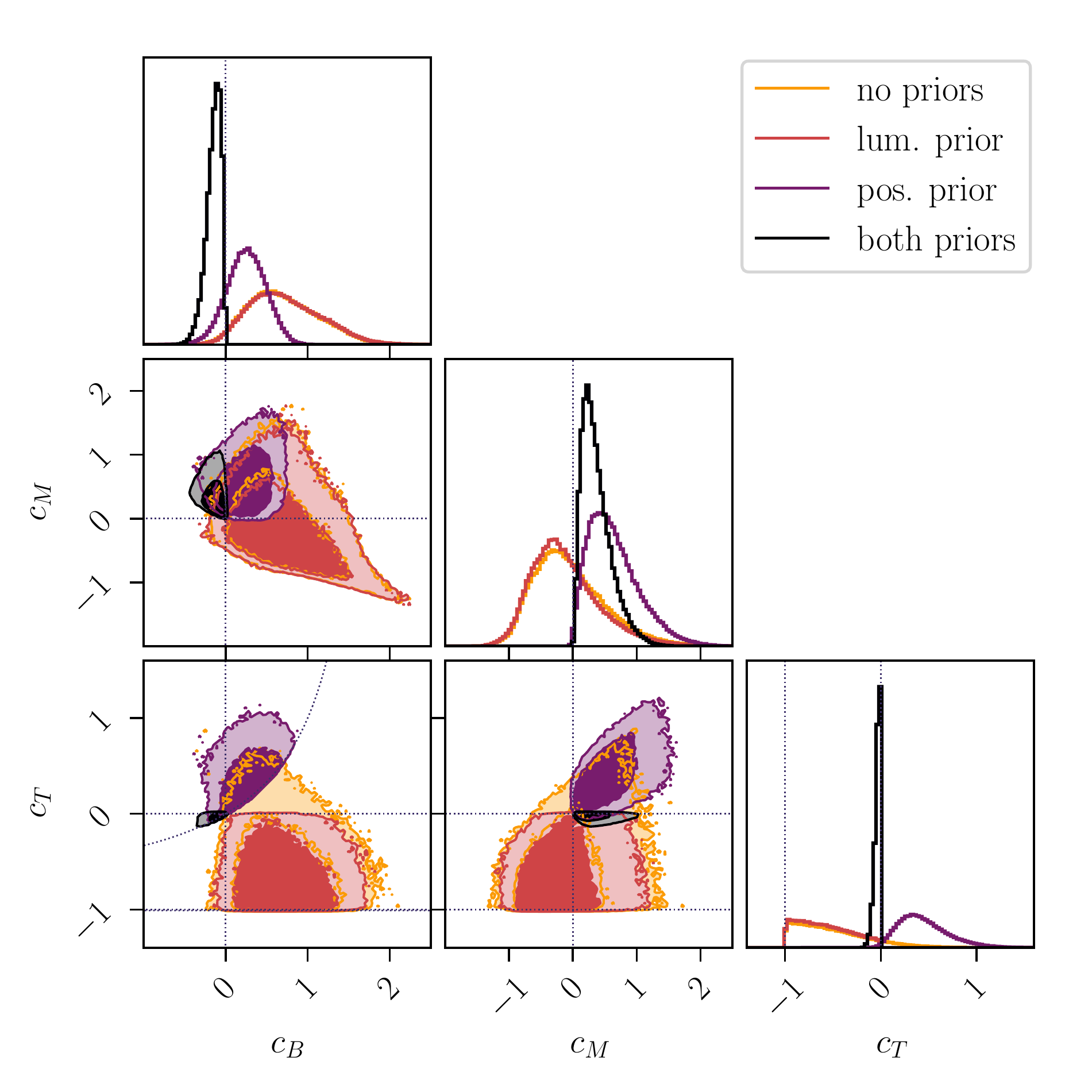}
\end{center}
\caption{Cosmological parameter constraints for the quartic Horndeski theory \eqref{quarticA}, using  $\aI = c_i \Omega_{\rm DE}$ \eqref{oParam} and different combinations of positivity \eqref{posprior} and (sub-)luminality priors \eqref{lumprior}. The positivity priors are derived from $\phi\phi \to \phi\phi$ scattering. Contours mark $68\%$ and $95\%$ confidence intervals, computed using CMB, RSD, BAO and matter power spectrum measurements. 
Dotted lines mark $c_i=0$ (the GR value), $c_T \geq -1$ (real GW speed) and $c_B < 2 c_T/(1+c_T)$ (positivity). 
The positivity prior eliminates over $60\%$ of the $2\sigma$ parameter space. If also combined with a (sub-)luminality prior, only $\lesssim 1\%$ of the $2\sigma$ parameter space survives.  
\label{fig1}}
\end{figure}

Expanding \eqref{quarticA} about a flat background ($g_{\mu \nu} = \eta_{\mu \nu} + h_{\mu \nu} /\MPl$), the tree-level scattering amplitude for $\phi \phi \to \phi \phi$ takes the form \eqref{Ast}, with, 
\begin{align}
%c_{ss} = \tfrac{1}{2} \bar{G}_{2, XX}  +  \bar{G}_{4,X}  , \;\,  c_{sst} = - \tfrac{3}{2} \left(  %\bar{G}_{4,XX} + \bar{G}_{4,X}^2 \right) 
c_{sst} &= - 6 \left(  \bar{G}_{4,XX} + \bar{G}_{4,X}^2  /  \G_4 \right), \nn \\ c_{ss} &=  2 \bar{G}_{2, XX}  +  4 \bar{G}_{4,X}  \G_{2,X} /  \G_4 ,
\end{align}
where an overbar indicates that the function is evaluated on the flat background ($\langle \phi \rangle = 0$).
From \eqref{cABposbounds}, the existence of a UV completion therefore requires
\begin{align} \label{posbounds}
2 \frac{\bar G_{4,X}}{\vphantom{\hat{G}}\bar{G}_4} &\geq -  \frac{\bar{G}_{2,XX}}{\G_{2,X}}  \,  , &   \frac{\bar{G}_{4,X}^2}{\vphantom{\hat{G}}\bar{G}_4}  &\leq - \bar G_{4,XX}  \, ,  
\end{align}
where we have assumed $\Lambda_2 \gg \Lambda_3$.
The other elastic amplitudes, $\phi h \to \phi h$ and $h h \to h h$, vanish at this order in $1/M_P$, and so scattering with external gravitons does not impose any additional positivity constraints.  
We show the analogous bounds for a general Horndeski theory \eqref{Horn_action} in the Appendix.
The above amplitudes and corresponding positivity bounds have been derived on a flat background. However, since \eqref{quarticA} is fully covariant, we may also consider the evolution of fluctuations about a cosmological background (i.e. a $\Lambda${}CDM background), and can assume that the positivity bounds \eqref{posbounds} continue to hold for the $G_i$ evaluated on the cosmological $\langle \phi \rangle$.   
\\

\noindent \textit{Linear perturbations in cosmology:} 
Cosmological deviations from GR are especially tightly constrained at the level of linear perturbations. We will therefore follow the approach of Refs. \cite{BelliniParam,Alonso:2016suf}, assuming a $\Lambda${}CDM-like background (motivated by the observed proximity to such a solution) and constraining perturbations around it. When perturbing \eqref{Horn_action} (c.f. \cite{Kobayashi:2011nu}), one finds that three independent combinations of the $G_i$ control the linear phenomenology \cite{Bellini:2014fua}: $\aM$, the running of the effective Planck mass $M_{\rm Pl}^{\rm eff} \equiv M \MPl$; $\aB$, the ``braiding'' that quantifies kinetic mixing between the metric and scalar perturbations; and $\aT$, the tensor speed excess, related to the sound speed of tensor perturbations via $c_{GW}^2 = 1 + \aT$. 
A fourth independent combination, the kineticity $\aK$, is effectively unconstrained at the level of linear perturbations and does not affect constraints on other parameters \cite{BelliniParam,Alonso:2016suf} (we have explicitly verified this in the present context), so we will not discuss it here. 
For the general Horndeski theory \eqref{Horn_action}  the $\aI$ are given in the Appendix.
For our specific example \eqref{quarticA}, one finds
%the $\aI$ are given by  
\begin{align}
%M^2 &= 2\left(G_4-2XG_{4,X}\right) , \nonumber \\
M^2\aM &=-2\frac{\dot X}{H}\left(G_{4,X} + 2 X G_{4,XX}\right), \nonumber \\
%HM^2\aB &= 2H M^2\aT+ 16X^2HG_{4,XX} \nonumber , \\ 
M^2\aB &= 8X\left(G_{4,X}+2XG_{4,XX}\right) \nonumber , \\ 
M^2\aT &= 4X G_{4,X} \,, 
\label{alpha2}
\end{align}
where $M^2 = 2\left(G_4-2XG_{4,X}\right)$. It will be instructive to re-arrange the expressions for $\aM$ and  $\aB$ and express them as 
\begin{align}
\aB &= 2\aT+ 16\frac{X^2}{M^2}G_{4,XX}, %\nonumber \\
&\aM &=-\frac{1}{4}\frac{\dot X}{HX}\aB,
\label{alpha3}
\end{align}
Having expressed the $\alpha_i$ in terms of the $G_i$ and their derivatives, we are now in a position to translate the positivity bounds into priors on the $\alpha_i$.
The $c_{ss}$ bound on $\bar{G}_{2,XX}$ is not particularly constraining at this level 
since none of the $\aI$ in \eqref{alpha2} depend on $G_2$ (only $\aK$ depends on this).
However, the $c_{sst}$ bound is highly constraining, since in an expanding universe it demands,
%\begin{align} \label{posprior}
%&{\rm pos. prior}: %&\frac{G_{4,X}^2}{\bar{G}_4} &\leq - G_{4,XX} &\Rightarrow& 
%
%&\aB &\leq 2 \aT - \frac{2\aT^2}{1+\aT}, 
%\end{align}
\begin{align} \label{posprior}
&{\rm pos. prior}: &\aB &\leq \frac{2 \aT}{1+\aT}, 
\end{align}
where we have used \eqref{alpha3} as well as $M^2 = 2 G_4/(1+\aT)$.
Given only the very basic assumption that the Horndeski EFT \eqref{quarticA} can be completed in a ``standard'' Wilsonian way at high energies, we have obtained a constraint \eqref{posprior} on the $\alpha_i$.     
Naturally, given further theoretical assumptions about the EFT and its underlying dynamics, there are further constraints that can be placed on the $\alpha_i$.
For instance, if gravitational waves were assumed to travel (sub-)luminally at low energies, then this would translate into the condition,
\begin{align} \label{lumprior}
&{\rm lum. prior}: 
% &G_{4,X} &\leq 0 &\Rightarrow& 
&\aT &\leq 0  \, . 
\end{align}
As another example, if the background evolution is driven by a $\langle \phi \rangle$ with certain properties (monotonicity, for instance), this also translates into possible conditions on the $\alpha_i$---we will return to this point later.   
Combining some or all of these different priors\footnote{For example, demanding both the subluminality condition \eqref{lumprior} and the positivity bounds \eqref{posbounds} requires a certain degree of self-interaction $\bar G_{2,XX} \geq 0$, which in turn has implications for how the background $\langle \phi \rangle$ evolves.  
} allows us to selectively carve out regions in ``theory space'' and fit data to only the corresponding low-energy parameter space. We stress that the positivity bounds are the most fundamental of such theoretical requirements, and hence captured the widest possible range of consistent UV models.        
\\

\begin{table}[t!]
\begin{center}
{\renewcommand{\arraystretch}{1.4}
\setlength{\tabcolsep}{0.1cm}
\begin{tabular}{|l||ccc|} \hline
{} & $c_B$ & $c_M$ & $c_T$\\ \hline\hline
no priors & $0.71\substack{+0.90 \\ -0.71}$ & $-0.02\substack{+1.32 \\ -0.89}$ & $ -1^* \le c_T < 0.25$\\ \hline
lum. prior & $0.73\substack{+0.91 \\ -0.72}$ & $-0.09\substack{+1.29 \\ -0.84}$ & $-1^* \le c_T \leq 0^*$ \\ \hline
pos. prior & $0.26\substack{+0.46 \\ -0.46}$ & $0.67\substack{+0.97 \\ -0.58}$ & $0.46\substack{+0.64 \\ -0.41}$\\ \hline
both priors & $-0.16\substack{+0.13 \\ -0.22}$ & $0.39\substack{+0.60 \\ -0.32}$ & $-0.10 < c_T \le 0^*$\\ \hline 
\end{tabular}
}
\end{center}
\caption{Posteriors on the dark energy/modified gravity $c_i$ parameters \eqref{oParam} for the quartic Horndeski theory \eqref{quarticA} as displayed in figure \ref{fig1}, i.e. following from different combinations of positivity \eqref{posprior} and (sub-)luminality priors \eqref{lumprior}. Uncertainties shown denote the $95 \%$ confidence level. The distribution for $c_T$ is typically strongly skewed. We therefore do not give a mean value in such cases and denote limit values due to prior boundaries (when there is an excellent fit to the data on that boundary) with an asterisk. 
}
\label{tab1}
\end{table}

\noindent {\textit{Cosmological parameter constraints:} We are now in a position to compute constraints on the $\aI$ (and hence on the deviations from GR they parameterise) using cosmological data. To do so, we will
perform a Markov chain Monte Carlo (MCMC) analysis, using Planck 2015 CMB temperature, CMB lensing and low-$\ell$ polarisation data \cite{Planck-Collaboration:2016af, Planck-Collaboration:2016aa, Planck-Collaboration:2016ae}, baryon acoustic oscillation (BAO) measurements from SDSS/BOSS \cite{Anderson:2014, Ross:2015}, constraints from the SDSS DR4 LRG matter power spectrum shape \cite{Tegmark:2006} and redshift space distortion (RSD) constraints from BOSS and 6dF \cite{Beutler:2012, Samushia:2014}.
Computing cosmological constraints requires choosing a parametrisation for the $\aI$. Numerous such parametrisations exist -- for a discussion of their relative merits see Refs. \cite{Bellini:2014fua,BelliniParam,Linder:2015rcz,Linder:2016wqw,Denissenya:2018mqs,Lombriser:2018olq,Gleyzes:2017kpi,Alonso:2016suf,mcmc}. 
Here we will pick arguably the most frequently used \cite{Bellini:2014fua}: 
\be \label{oParam}
\aI = c_i \Omega_{\rm DE}.
\ee 
This parameterises each $\aI$ in terms of just one constant parameter, $c_i$, and is known to very accurately capture the evolution of a wide sub-class of Horndeski theories \cite{Pujolas:2011he,Barreira:2014jha} (for further details and a comparison of results for different parametrisations see Ref. \cite{mcmc} and references therein).
When imposing priors, we require them to be satisfied at all times, i.e. dynamically throughout the evolution until today {\it as well as} at late times, when $\Omega_{\rm DE} \to 1$ on our $\Lambda$CDM-like background. In the context of \eqref{oParam}, this late time limit yields the strongest bounds on the $c_i$, given the priors on the $\alpha_i$.\footnote{We have checked that our data constraints are only marginally different when compared to only imposing priors until today, i.e. in the interval $\Omega_{\rm DE} \in [0,0.7]$.}

We now compute constraints on the modified gravity/dark energy parameters $c_{B}, c_M$ and $c_{T}$, marginalising over the standard $\Lambda{\rm CDM}$ parameters $\Omega_{\rm cdm}, \Omega_{\rm b}, \theta_s,A_s,n_s$ and $\tau_{\rm reio}$ -- for technical details regarding the MCMC implementation see \cite{mcmc}. The results are shown in Figure~\ref{fig1} and Table~\ref{tab1}. %A number of features deserve highlighting: 
For the Horndeski action \eqref{quarticA}, applying the positivity prior \eqref{posprior} reduces the overall volume in $c_i$ parameter-space by a factor $\gtrsim 3$, i.e. eliminates $\gtrsim 60\%$ of the previously allowed parameter space\footnote{By the ``volume'' in parameter space we mean the very simple measure $\Delta c_B \Delta c_M \Delta c_T$, where $\Delta c_i$ denotes the 95\% confidence interval for $c_i$ (note that this measure is not unique and many alternative measures exist).  For example, with no priors we have $\Delta c_B = 0.90 + 0.71 = 1.61$.}. To show how this interfaces with other theoretical restrictions one may impose, we also include the effects of the (sub-)luminality prior \eqref{lumprior} -- this eliminates all but $\lesssim 1\%$ of the parameter space, showing how drastically such bounds can improve constraints on deviations from GR\footnote{In the joint case, a corollary of these improved constraints is that the allowed parameter values are pulled significantly closer to their GR values.}. 
A key reason for this is that the positivity prior, (sub-)luminality prior and data constraints act in a highly complementary fashion: without any priors, the data prefer negative $c_T$ and positive $c_B$, a combination that is ruled out by the positivity prior. That prior + data instead prefer a positive, (super-)luminal $c_T$, which in turn is in tension with the sub-luminality requirement (which by itself is only mildly constraining -- see Table~\ref{tab1}). So jointly applying both priors drastically reduces the available parameter space.
 
Physically, the positivity prior corresponds to requiring a ``standard'' UV completion beyond $\Lambda_3$ (in the sense discussed above) for the scalar sector of the theory.
Assuming such a UV completion, positivity bounds do not only significantly tighten constraints, but importantly also shift them by $\gtrsim 1 \sigma$,  stressing the importance of incorporating such bounds into the data analysis. If the underlying physics were to also mandate a (sub-)luminality prior, this statement is further strengthened and the majority of the $2\sigma$ confidence region computed without such joint priors can then lie in a fundamentally \emph{unphysical} region of parameter space, as the $c_T-c_B$ plane in Figure~\ref{fig1} shows.

The data constraints themselves are primarily driven by Planck CMB data, RSD measurements and gradient instabilities. Here Planck data constrain the $c_i$ primarily due to the way they modify the (late) ISW effect, 
as accurately probed by large scales in the CMB TT power spectrum. 
Secondly, RSDs provide a complementary probe of galaxy clustering. This rules out large positive $c_M$, since this would lead to too much clustering (the rate of structure growth becomes too large in that case). 
Thirdly, gradient instabilities are associated with an imaginary ``speed of sound'', leading to a dangerous growth of perturbations. For scalar perturbations, such instabilities occur when
\begin{align} \label{gradient_condition}
(2-\aB)\left(\hat\alpha - \frac{\dot H}{H^2}\right) - \frac{3(\rho_{\rm tot} + p_{\rm tot})}{H^2M^2} + \frac{\dot{\alpha}_B}{H} < 0,
\end{align}
where $\rho_{\rm tot}$ and $p_{\rm tot}$ are the total energy density and pressure in the universe and we have defined $\hat \alpha \equiv \tfrac{1}{2}\aB(1+\aT) + \aM - \aT$. %
The onset of these instabilities rules out large negative $\alpha_M$ and $\alpha_B$. For tensor perturbations, the analogous constraint simply imposes $\alpha_T \geq -1$. 
Note that we do not rule out solutions with gradient instabilities {\it a priori}, but find that the data rule out solutions with significant such instabilities by themselves. Also note that, when using both priors, including RSD measurements no longer has a significant effect, since the relevant parts of parameter space are ruled out by the priors already. 

Finally, from \eqref{alpha3} and Figure~\ref{fig1} one can observe that additional priors on the background evolution for $\phi$ (which we have remained agnostic about here) have the potential to rule out the simple EFT \eqref{quarticA} altogether. For instance, assuming both priors \eqref{posprior} and \eqref{lumprior}, $\aM$ can only be positive if $\dot X > 0$.
Indeed this illustrates a more general point: If one has information about the full covariant theory, additional information e.g. from the background evolution of the field 
can be used to place further constraints on the theory. For such cases, while the current state of the art of Einstein-Boltzmann solvers does not allow this yet \cite{Hu:2013twa,Zumalacarregui:2016pph}, implementing the full background evolution of the fields into the present analysis would therefore be a highly promising avenue for the future.
%
%\\
\\

\noindent\textit{Conclusions:} We have developed a holistic approach to deriving cosmological parameter constraints on deviations from GR while simultaneously taking into account {\it both} ``positivity'' priors from fundamental physics {\it and} constraints from current observational data.
In doing so, we have computed and discussed new positivity bounds for the general class of Horndeski theories. 
Using a particularly simple subclass of these theories as an example, we have explicitly shown that merging these bounds with current data can significantly improve constraints on deviations from GR, eliminating $\gtrsim 60\%$ of the previously allowed parameter space. We have also shown that combining fundamental positivity requirements with further theoretical priors can drastically improve constraints, for instance an additional sub-luminality prior for the speed of gravitational waves eliminates all but $\lesssim 1\%$ of the previously allowed parameter space. 
To place this improvement in a broader context, constraints from future CMB S-4, LSST and SKA data are forecast \cite{Alonso:2016suf} to shrink the currently allowed parameter space by a factor of $\sim 20$ (to be compared\footnote{Encouragingly, these forecasts indicate that $c_M$, the least well constrained parameter in all cases considered here, will be the one most tightly constrained by future data. This suggests that the fruitful complementarity between the theoretical bounds considered here and pure data constraints will persist into the (near-)future.} with the factor $\sim 3$ improvement from using positivity priors and the factor $\sim 140$ improvement from combining positivity and sub-luminality priors).
Another example of future data that promise to strongly constrain such cosmological theories is future gravitational wave speed measurements at frequencies firmly within the regime of validity for such theories \cite{deRham:2018red} -- measurements that e.g. have the potential to rule out most of the simple subclass of theories we have focused on here.
More generally, should future data collectively pull the contours into the ``positivity'' region, this will experimentally confirm the QFT nature of the underlying UV physics, and the priors presented here will have allowed for significantly improved parameter estimation in advance of that future data. Conversely, should there be increased tension between future observations and the positivity bounds, this is evidence that our Universe does \emph{not} resemble a standard QFT with a Lorentz-invariant vacuum, providing a qualitatively new probe of the high energy regime.
Finally we stress that general Horndeski models are currently not constrained as strongly as the example subclass we have focused on (see the Appendix for details), yet the example given clearly illustrates the strong potential constraining power of positivity bounds.
Indeed, with several additional positivity bounds expected to exist (from going beyond tree-level $2 \to 2$ scattering on flat space), this underlines how essential and promising a joint approach merging fundamental physical priors with data constraints will be in going forward.
In order to maximally constrain deviations from GR using future data, it will be key to ensure one is working with a physical parameter space (instead of overfitting the data with unphysical parameter choices) along the lines outlined here.
\\

%\vspace{-0.1in}
\section*{Acknowledgments}
\vspace{-0.1in}
\noindent We thank D. Alonso, E. Bellini, C. de Rham, P. Ferreira, L. Santoni, A. Tolley and E. Trincherini for useful discussions and comments on a draft. JN acknowledges support from Dr. Max R\"ossler, the Walter Haefner Foundation and the ETH Zurich Foundation. 
SM is supported by an Emmanuel College Research Fellowship and partially supported by STFC consolidated grant ST/P000681/1.  
In deriving the results of this paper, we have used: CLASS \cite{Blas:2011rf},  corner \cite{corner}, hi\_class \cite{Zumalacarregui:2016pph}, MontePyton \cite{Audren:2012wb,Brinckmann:2018cvx}, xAct \cite{xAct} and xIST \cite{xIST}.

%%%%%%%%
\appendix
\section*{Appendix}
%%%%%%%%
% Change equation numbers from xx to A.xx so that they don't clash with main file 
%\renewcommand{\theequation}{A.\arabic{equation}}

\noindent Here we give positivity bounds on the general Horndeski theory \eqref{Horn_action}\footnote{
The form of the interactions we use was rediscovered independently by \cite{Deffayet:2011gz}, and later shown to be equivalent to the original Horndeski theory \cite{Kobayashi:2011nu}.  
} and briefly discuss how these will link with constraints derived from linear cosmology. Indices are raised and lowered with the full metric $g_{\mu\nu}$ and we employ a $(-+++)$ signature.
\\

\noindent {\it Positivity bounds}: 
The $\phi \phi \to \phi \phi$ amplitudes can be written in the form \eqref{Ast}, with\footnote{
There is also an overall factor of $\bar{G}_{2,X}$ from the wavefunction normalisation of $\phi$, which we do not show here, since we will assume that $\bar{G}_{2,X}$ is positive in order to avoid ghost instabilities around flat backgrounds and therefore this overall factor will not affect any positivity bound here. 
}, 
\begin{align}
 c_{ss}   &=   2 \left(  \bar{G}_{2,XX} -  2 \bar{G}_{4,\phi \phi X} \right)
 - 2 \frac{ \bar G_{2, \phi \phi} }{ \G_{2,X} }  \left(   \bar{G}_{3,X} + 3 \bar{G}_{4, \phi X}      \right)^{2}
    \nn \\
 &\quad  + 4  ( \bar{G}_{2, \phi X} +  \bar{G}_{3, \phi \phi} ) ( \bar{G}_{3,X} + 3 \bar{G}_{4, \phi X} ) \,   \nn  \\
&\quad +   \frac{2}{\vphantom{\hat{G}}\G_4}    ( 2 \G_{4,X} -  \G_{5,\phi} ) ( \G_{2,X} + \G_{3,\phi} - \G_{4,\phi\phi} )   \, ,   \label{eqn:A1} \\
 c_{sst}   &=   - 2  \left( 3 \bar{G}_{4, XX}  -  \bar{G}_{5, \phi X} \right) 
+ 3 \left(  \bar{G}_{3,X} + 3 \bar{G}_{4, \phi X}   \right)^2  \nn  \\
&\quad - \frac{3}{2 \vphantom{\hat{G}}\bar{G}_4 } (  2 \bar{G}_{4,X} -  \G_{5,\phi}    )^2     \,  ,  \label{eqn:A2}   
\end{align}
which we derive at the end of this Appendix. 
The overbar denotes the function evaluated at $\phi = 0$ and the mass of the scalar is given by $\bar G_{2, \phi \phi}/\bar{G}_{2,X} = -m^2/H_0^2$.
These obey the positivity bounds \eqref{cABposbounds}, i.e.
\begin{align} \label{posbounds_full}
 c_{ss}  &\geq 0,  &c_{sst}  &\geq 0 \, ,
\end{align}
again assuming that $\Lambda_2 \gg \Lambda_3$. 
Imposing a shift symmetry ($\phi \to \phi + c$, where $c$ is a constant) gives the simplified bounds, 
\begin{equation}
\label{posbounds_shift}
2 \G_{2,X} \bar{G}_{4,X}  \geq - \bar{G}_{2,XX} \G_4 , \;\;\;\;  2 \bar{G}_{4,XX} + 2 \bar{G}_{4,X}^2/ \G_4  \leq  \bar{G}_{3,X}^2  \,  \, ,
\end{equation}
since then the $G_n$ are functions of $X$ only\footnote{
Since $G_3$ and $G_5$ multiply total derivatives, a non-zero $G_{3,\phi}$ and $G_{5,\phi}$ would be compatible with a shift symmetry---however these can be absorbed into $G_{2,X}$ and $G_{4,X}$ up to a total derivative. 
}. 

As a consistency check, note that in the shift-symmetric case one can expand the Horndeski terms on flat space as simply the cubic and quartic Galileon,
\begin{align}
\mathcal{L}_3 &= \frac{g_3}{3! \Lambda_3^3} \phi \left( [ \hat  \Phi]^2 - [ \hat \Phi^2]   \right)  + ... \; , \\
\mathcal{L}_4 &= \frac{g_4}{4! \Lambda_3^6} \phi \left(   [ \hat \Phi]^3 - 3 [ \hat \Phi ] [ \hat \Phi^2] + 2 [ \hat \Phi^3 ]         \right) + ... \; ,
\end{align}
where $\hat \Phi^\mu_{\;\; \nu} = \partial^\mu \partial_\nu \phi$,  $\bar G_{4,XX}/4 = g_4/4!$ and $\bar G_{3,X}/3 = g_3/3!$, to use the notation of Ref. \cite{deRham:2017imi}. There, the forward limit positivity bound $4 g_4 \leq 3 g_3^2$ was found\footnote{
A similar bound was also observed at $t\neq0$ in \cite{Nicolis:2009qm}. 
}, which is consistent with \eqref{posbounds_shift}. A stronger bound was also obtained in that case by demanding that the effects of new physics come in at a scale which is parametrically larger than $m^2$, the mass of $\phi$, whereas here we have simply demanded that new physics can enter at \emph{any} scale before $\Lambda_3$ to restore unitarity in the EFT.  
\\
\\

\noindent {\it Positivity caveats}: 
Strictly speaking, the positivity bounds \eqref{cABposbounds} were established assuming (i) that all particles involved have a nonzero mass, and (ii) a flat background (i,e. trivial vacuum expectation values) for the fields.
The mass is technically important in that it connects polynomial boundedness to locality via the Froissart bound, ensures no divergent $t$-channel pole in the forward limit, and guarantees an analytic Mandelstam triangle, $0< s, t < 4m^2$.
Here, however, we have also applied these constraints in the presence of a (massless) graviton. We do this with the understanding that we always work to leading order in $\MPl$ and treat all gravitational effects semiclassically. For instance, beyond the leading order $1/\Lambda_2^4$ contribution to the amplitudes there is a $t$-channel pole from virtual graviton exchange, $\sim s^2/(\MPl^2 t)$, which we have neglected\footnote{
See \cite{Bellazzini:2019xts} for a recent, more sophisticated, proposal for remedying the formal divergence as $t \to 0$ in the $\mathcal{A}$.
} (since, although formally divergent in the forward limit, $t \to 0$, it vanishes in the limit $\MPl \to \infty$).  
We have also assumed that the bounds continue to hold (at least approximately) on a cosmological background. This is particularly well-motivated in cases where the two vacua (flat and cosmological) are connected by a smooth limit -- for instance by taking $H \to 0$ and $\dot H \to 0$ while sending $\MPl \to \infty$, as described by Ref. \cite{Baumann:2015nta} in the context of the EFT of inflation.
In the future, the positivity bounds \eqref{cABposbounds} are likely to be further improved by better exploiting the analytic structure of massless particles and de Sitter isometries. 
\\

\noindent {\it Linear cosmology}: For the general Horndeski theory \eqref{Horn_action}, the $\aI$ controlling the dynamics of linear perturbations (around an FRW background) are given by
\begin{align}
M^2 &= 2\left(G_4-2XG_{4,X}+XG_{5,\phi}-\frac{{\dot \phi}H}{H_0^2} XG_{5,X}\right) , \nonumber \\
M^2\aB &= - 2\frac{\dot{\phi}}{H}\left(XG_{3,X}+G_{4,\phi}+2XG_{4,\phi X}\right) \nonumber \\ & 
+8X\left(G_{4,X}+2XG_{4,XX}-G_{5,\phi}-XG_{5,\phi X}\right) \nonumber \\ & 
+2\frac{\dot{\phi}H}{H_0^2}X\left(3G_{5,X}+2XG_{5,XX}\right) \nonumber , \\ 
M^2\aT &= 2X\left[2G_{4,X}-2G_{5,\phi}-\left(\frac{\ddot{\phi}}{H_0^2}-\frac{\dot{\phi}H}{H_0^2}\right)G_{5,X}\right] \,,
\label{alphadef}
\end{align}
where we have defined $HM^2\aM \equiv \tfrac{d}{dt}M^2$ and omitted $\aK$, as before.\footnote{Note that there is a (conventional) sign difference for $G_3$ in the expressions for the $\alpha_i$ here compared to \cite{Bellini:2014fua}. This is due to a sign difference in our formulation of the Horndeski action \eqref{Horn_action} compared to the corresponding formulation of \cite{Bellini:2014fua}. Factor of $H_0$ differences are due to our dimensionless definition of the $G_i$ (as opposed to the dimensionful $G_i$ in \cite{Bellini:2014fua}).} Imposing shift symmetry and eliminating $G_3$ and $G_5$ then recovers \eqref{alpha2}.
Note that shift-symmetric Horndeski theories have a number of attractive properties related to radiative stability \cite{Pirtskhalava:2015nla} (also see Ref. \cite{radstab}). Finally, note that the scalar speed of sound, $c_s$, satisfies \cite{Kobayashi:2011nu,Bellini:2014fua,Zumalacarregui:2016pph,mcmc}
\be \label{cs_full}
{\cal D} c_s^2 = (2-\alpha_B)\left(\hat\alpha - \frac{\dot H}{H^2}\right) - \frac{3(\rho_{\rm tot} + p_{\rm tot})}{H^2M^2} + \frac{\dot \alpha_B}{H},
\ee
where ${\cal D} \equiv \alpha_K + \tfrac{3}{2}\alpha_B^2$ (which is positive in the absence of ghost-like instabilities) and we recall that $\hat \alpha \equiv \tfrac{1}{2}\aB(1+\aT) + \aM - \aT$. This shows that requiring (sub-)luminality for $c_s$ is an orthogonal constraint to the positivity bounds considered here. More specifically, given a choice of otherwise consistent $\alpha_i$, such a constraint here simply places a lower bound on $\alpha_K$. 
\\

\noindent {\it Cosmological parameter constraints}: For fully general Horndeski scalar-tensor theories, the constraining power of the {\it current} positivity bounds is therefore clearly limited: there are currently too few bounds \eqref{posbounds_full} to strongly constrain the freedom in the general $\alpha_i$ \eqref{alphadef}. In contrast, for the specific subclass of models \eqref{quarticA} we focused on in the main text, the positivity bounds could straightforwardly be expressed in terms of a constraint on the $\alpha_i$.
Analogously, the bounds \eqref{posbounds_full} can already be highly constraining for other specific `simple enough' Horndeski models, i.e. specific choices of the $G_i$ with a relatively small number of free parameters.
In order to achieve a similar level of constraining power for more general models, additional theoretical constraints will be needed. Fortunately the positivity bounds discussed here are only the beginning, since additional bounds are expected to arise from going beyond $2 \to 2$ tree-level scattering on flat space, These will complement constraints from the wealth of upcoming, near-future observational data, so fully exploring the impact of such fundamental theoretical requirements on cosmological parameter estimation will be of crucial importance going forward.  
\\

\noindent {\it Computing the amplitude}: 
Finally, let us detail the computation of the 2-to-2 amplitudes in the Horndeski theory. 
We will focus on the interactions which can contribute to elastic scattering, and will count in powers of large $M_P$ (treating $\Lambda_3$ as fixed). \\

Neglecting all total derivatives, the leading order vertices from the Horndeski Lagrangian are, \\
\FloatBarrier
\begin{figure}[h!]
\begin{flushleft}
		\begin{tikzpicture}[baseline=-0.6cm]
			\begin{feynman}
				\vertex (a1);
				\vertex [below=0.5cm of a1] (b1);
				\vertex [below=0.5cm of b1] (c1);
				\vertex [left=0.5cm of a1] (a2);
				\vertex [below=0.5cm of a2] (b2);
				\vertex [below=0.5cm of b2] (c2);
				\vertex [left=0.5cm of a2] (a3);
				\vertex [below=0.5cm of a3] (b3);
				\vertex [below=0.5cm of b3] (c3);
				\vertex [left=0.5cm of a3] (a4);
				\vertex [below=0.5cm of a4] (b4);
				\vertex [below=0.5cm of b4] (c4);
				
				\diagram*{
				(a3) -- [scalar] (b2),
				(c3) -- [scalar] (b2),
				(a1) -- [scalar] (b2),
				(c1) -- [scalar] (b2),
				};				
			\end{feynman}		
		\end{tikzpicture}$=  \frac{ - 3 \G_{4,XX}  + \G_{5,\phi X}  }{ 3 \Lambda_3^6} \; \delta^{\mu \alpha \rho}_{\nu \beta \sigma} \; \phi \phi_{\mu}^{\nu} \phi_{\alpha}^{\beta} \phi_{\rho}^{\sigma}  $  \\[5pt]

		\begin{tikzpicture}[baseline=-0.6cm]
			\begin{feynman}
				\vertex (a1);
				\vertex [below=0.5cm of a1] (b1);
				\vertex [below=0.5cm of b1] (c1);
				\vertex [left=0.5cm of a1] (a2);
				\vertex [below=0.5cm of a2] (b2);
				\vertex [below=0.5cm of b2] (c2);
				\vertex [left=0.5cm of a2] (a3);
				\vertex [below=0.5cm of a3] (b3);
				\vertex [below=0.5cm of b3] (c3);
				\vertex [left=0.5cm of a3] (a4);
				\vertex [below=0.5cm of a4] (b4);
				\vertex [below=0.5cm of b4] (c4);
				
				\diagram*{
				(b3)  -- [scalar] (b2),
				(b2) -- [scalar] (a1),
				(b2) -- [scalar] (c1),
				};				
			\end{feynman}		
		\end{tikzpicture} $= - \frac{   \G_{3,X} + 3 \G_{4,\phi X}   }{3 \Lambda_3^3} \; \delta^{\mu \nu}_{\alpha \beta} \; \phi \phi_\mu^{\nu} \phi_\alpha^{\beta}  $  \\[5pt]

		\begin{tikzpicture}[baseline=-0.6cm]
			\begin{feynman}
				\vertex (a1);
				\vertex [below=0.5cm of a1] (b1);
				\vertex [below=0.5cm of b1] (c1);
				\vertex [left=0.5cm of a1] (a2);
				\vertex [below=0.5cm of a2] (b2);
				\vertex [below=0.5cm of b2] (c2);
				\vertex [left=0.5cm of a2] (a3);
				\vertex [below=0.5cm of a3] (b3);
				\vertex [below=0.5cm of b3] (c3);
				\vertex [left=0.5cm of a3] (a4);
				\vertex [below=0.5cm of a4] (b4);
				\vertex [below=0.5cm of b4] (c4);
				
				\diagram*{
				(b3)  -- [photon] (b2),
				(b2) -- [scalar] (a1),
				(b2) -- [scalar] (c1),
				};				
			\end{feynman}		
		\end{tikzpicture} $=  \frac{ 2 \G_{4,X} -  \G_{5,\phi}  }{2 \Lambda_3^3} \; \delta^{\mu \alpha\rho}_{\nu\beta\sigma} \; \phi  \phi_\mu^\nu  ( h^{\beta}_{\alpha} )_{,\rho}^{,\sigma}  $
\end{flushleft}
\end{figure} 
\FloatBarrier
%%%%%%%%%%%%%%%%%%%%%%%%%%
\noindent and the subleading vertices are, \vspace{-0.1cm}
%%%%%%%%%%%%%%%%%%%%%%%%%%
\FloatBarrier
\begin{figure}[htbp!]
\begin{flushleft}
		\begin{tikzpicture}[baseline=-0.6cm]
			\begin{feynman}
				\vertex (a1);
				\vertex [below=0.5cm of a1] (b1);
				\vertex [below=0.5cm of b1] (c1);
				\vertex [left=0.5cm of a1] (a2);
				\vertex [below=0.5cm of a2] (b2);
				\node [below=0.425cm of a2, dot] (d);
				\vertex [below=0.5cm of b2] (c2);
				\vertex [left=0.5cm of a2] (a3);
				\vertex [below=0.5cm of a3] (b3);
				\vertex [below=0.5cm of b3] (c3);
				\vertex [left=0.5cm of a3] (a4);
				\vertex [below=0.5cm of a4] (b4);
				\vertex [below=0.5cm of b4] (c4);
				
				\diagram*{
				(a3) -- [scalar] (b2),
				(c3) -- [scalar] (b2),
				(a1) -- [scalar] (b2),
				(c1) -- [scalar] (b2),
				};				
			\end{feynman}		
		\end{tikzpicture}  $=   \frac{ 3 \G_{2,XX} + 2 \G_{3, \phi X}  }{6 M_P \Lambda_3^3}  \phi_\mu \phi^\mu \phi_\nu \phi^\nu   $   \\
$\;$ ~ \qquad\qquad		$ - \frac{ \G_{3,\phi X}  +  3  \G_{4,\phi\phi X} }{6 M_P \Lambda_3^3} \; \delta^{\mu \alpha}_{\nu \beta} \; \phi \phi \phi_{\mu}^\nu \phi_\alpha^\beta $ \\[5pt]
		
		\begin{tikzpicture}[baseline=-0.6cm]
			\begin{feynman}
				\vertex (a1);
				\vertex [below=0.5cm of a1] (b1);
				\vertex [below=0.5cm of b1] (c1);
				\vertex [left=0.5cm of a1] (a2);
				\vertex [below=0.5cm of a2] (b2);
				\node [below=0.425cm of a2, dot] (d);
				\vertex [below=0.5cm of b2] (c2);
				\vertex [left=0.5cm of a2] (a3);
				\vertex [below=0.5cm of a3] (b3);
				\vertex [below=0.5cm of b3] (c3);
				\vertex [left=0.5cm of a3] (a4);
				\vertex [below=0.5cm of a4] (b4);
				\vertex [below=0.5cm of b4] (c4);
				
				\diagram*{
				(a3) -- [scalar] (b2),
				(c3) -- [photon] (b2),
				(a1) -- [scalar] (b2),
				(c1) -- [photon] (b2),
				};				
			\end{feynman}		
		\end{tikzpicture}$= {\color{blue} \frac{ - 2 \G_{4,X}  +  \G_{5,\phi} }{2 M_P \Lambda_3^3}  \phi \phi_{\mu}^{\nu}  \Big(  2 \delta^{\mu\alpha\rho\kappa}_{\nu\beta\sigma \iota} h_\alpha^{\beta} ( h_{\rho}^{\sigma} )^{, \iota}_{, \kappa} }$ \\
~		\qquad\qquad\qquad\qquad\qquad ${\color{blue} - \delta^{\alpha \rho}_{\beta\sigma} ( h^\mu_{\alpha} )_{,\rho}   ( h^\beta_\nu  )^{, \sigma}  + ( h_{\alpha}^{\beta} )_{,\mu} ( h^{\alpha}_{\beta}  )^{,\nu}    \Big) }$  \hspace{1.5cm} \hskip 10pt\\[5pt]
		\begin{tikzpicture}[baseline=-0.6cm]
			\begin{feynman}
				\vertex (a1);
				\vertex [below=0.5cm of a1] (b1);
				\vertex [below=0.5cm of b1] (c1);
				\vertex [left=0.5cm of a1] (a2);
				\vertex [below=0.5cm of a2] (b2);
				\node [below=0.425cm of a2, dot] (d);
				\vertex [below=0.5cm of b2] (c2);
				\vertex [left=0.5cm of a2] (a3);
				\vertex [below=0.5cm of a3] (b3);
				\vertex [below=0.5cm of b3] (c3);
				\vertex [left=0.5cm of a3] (a4);
				\vertex [below=0.5cm of a4] (b4);
				\vertex [below=0.5cm of b4] (c4);
				
				\diagram*{
				(b3)  -- [scalar] (b2),
				(b2) -- [scalar] (a1),
				(b2) -- [scalar] (c1),
				};				
			\end{feynman}		
		\end{tikzpicture} $=  \frac{ \G_{2,\phi X } + \G_{3, \phi \phi} }{2 M_P} \, \, \phi \phi \phi_\mu^\mu  $  \hspace{1.5cm} \hskip 10pt\\[5pt]

		\begin{tikzpicture}[baseline=-0.6cm]
			\begin{feynman}
				\vertex (a1);
				\vertex [below=0.5cm of a1] (b1);
				\vertex [below=0.5cm of b1] (c1);
				\vertex [left=0.5cm of a1] (a2);
				\vertex [below=0.5cm of a2] (b2);
				\node [below=0.425cm of a2, dot] (d);
				\vertex [below=0.5cm of b2] (c2);
				\vertex [left=0.5cm of a2] (a3);
				\vertex [below=0.5cm of a3] (b3);
				\vertex [below=0.5cm of b3] (c3);
				\vertex [left=0.5cm of a3] (a4);
				\vertex [below=0.5cm of a4] (b4);
				\vertex [below=0.5cm of b4] (c4);
				
				\diagram*{
				(b3)  -- [photon] (b2),
				(b2) -- [scalar] (a1),
				(b2) -- [scalar] (c1),
				};				
			\end{feynman}		
		\end{tikzpicture} $=  \frac{ \G_{2,X} + \G_{3,\phi}  }{M_P} h_{\mu\nu} \left(  \phi^\mu \phi^\nu - \frac{1}{2} \eta^{\mu\nu} \phi^\rho \phi_\rho     \right)   $ \\
$\;$~\qquad\qquad $	+ \frac{\G_{4,\phi\phi}}{4M_P} \; \delta_{\nu\beta}^{\mu \alpha} \; \phi \phi (h_{\mu}^{\nu}  )_{,\alpha}^{,\beta}  $   \\[5pt]

		\begin{tikzpicture}[baseline=-0.6cm]
			\begin{feynman}
				\vertex (a1);
				\vertex [below=0.5cm of a1] (b1);
				\vertex [below=0.5cm of b1] (c1);
				\vertex [left=0.5cm of a1] (a2);
				\vertex [below=0.5cm of a2] (b2);
				\node [below=0.425cm of a2, dot] (d);
				\vertex [below=0.5cm of b2] (c2);
				\vertex [left=0.5cm of a2] (a3);
				\vertex [below=0.5cm of a3] (b3);
				\vertex [below=0.5cm of b3] (c3);
				\vertex [left=0.5cm of a3] (a4);
				\vertex [below=0.5cm of a4] (b4);
				\vertex [below=0.5cm of b4] (c4);
				
				\diagram*{
				(b3)  -- [scalar] (b2),
				(b2) -- [photon] (a1),
				(b2) -- [photon] (c1),
				};				
			\end{feynman}		
		\end{tikzpicture} $= {\color{blue} - \frac{ \G_{4, \phi} }{4 M_P} \,\left(  \delta^{\mu \alpha}_{\nu \beta}  \; \phi_\mu^{\nu}  h_\alpha^\rho h_{\rho}^{\beta}  + \phi \, \left( h_{\alpha \beta} \right)_{,\mu} \left( h^{\alpha \beta} \right)^{, \mu}    \right) }  $  \hspace{1.5cm} \hskip 10pt\\[5pt]

		\begin{tikzpicture}[baseline=-0.6cm]
			\begin{feynman}
				\vertex (a1);
				\vertex [below=0.5cm of a1] (b1);
				\vertex [below=0.5cm of b1] (c1);
				\vertex [left=0.5cm of a1] (a2);
				\vertex [below=0.5cm of a2] (b2);
				\node [below=0.425cm of a2, dot] (d);
				\vertex [below=0.5cm of b2] (c2);
				\vertex [left=0.5cm of a2] (a3);
				\vertex [below=0.5cm of a3] (b3);
				\vertex [below=0.5cm of b3] (c3);
				\vertex [left=0.5cm of a3] (a4);
				\vertex [below=0.5cm of a4] (b4);
				\vertex [below=0.5cm of b4] (c4);
				
				\diagram*{
				(b3)  -- [photon] (b2),
				(b2) -- [photon] (a1),
				(b2) -- [photon] (c1),
				};				
			\end{feynman}		
		\end{tikzpicture} $= \frac{ \G_{4} }{ M_P} \, \delta^{3} \left( \sqrt{-g} R  \right) $
		\hspace{1.5cm} \hskip 10pt 
\end{flushleft}
\end{figure} \FloatBarrier  \vspace{-0.3cm}
\noindent where the blue expressions for the subleading $\phi \phi hh$ and $\phi hh$ vertices have assumed that both gravitons are on-shell (transverse and traceless). The $hhh$ vertex is the usual cubic interaction of GR. 
 
The scalar propagator, 
\begin{equation}
\phi
\; \feynmandiagram [horizontal=a to b] {
a -- [scalar] b
}; \;
\phi =  \mathcal{P} ( p ) 
\end{equation}
is given in momentum space by,
\begin{equation}
 \mathcal{P} (p) p^2 = -i 
\end{equation}
and similarly the graviton propagator,
\begin{equation}
h_\mu^\nu
\; \feynmandiagram [horizontal=a to b] {
a -- [photon] b
}; \;
h_\alpha^\beta =  \mathcal{P}_{\mu \alpha}^{\nu \beta} ( p ) 
\end{equation}
is defined by the relation, 
\begin{equation}
\frac{\G_4}{2} \mathcal{P}_{\mu \alpha}^{\nu \beta} (p) \, \delta_{\beta \sigma \nu'}^{\alpha \rho \mu'} p_{\rho} p^{\sigma}  = -i \delta_{\mu}^{\mu'} \delta_{\nu}^{\nu'}  \, .
\label{eqn:hprop}
\end{equation}
%We will set $\G_4 = 2$ for a canonically normalized propagator. 
The total amplitudes are given by the Feynman diagrams shown in Figure \ref{fig:feyn1} and \ref{fig:feyn2}. The $\phi \phi \to \phi \phi$ amplitudes are written explicitly in \eqref{eqn:A1} and \eqref{eqn:A2}, while the $\phi h \to \phi h$ amplitude is found to exactly vanish at this order (in particular, note that the leading $\phi \phi h$ vertex contains precisely the same structure as \eqref{eqn:hprop} and so there is no graviton pole at this order). This is not surprising from the point of view of field redefinitions (which preserves the scattering amplitudes)---one can always transform to the Einstein frame perturbatively, 
\begin{align}
 h_{\mu\nu} \to h_{\mu\nu} + \frac{ 2 \G_{4,X} - \G_{5,\phi} }{\vphantom{\hat{G}}\G_4 \Lambda_2^4} \phi_\mu \phi_\nu
\end{align}
which removes the leading order $h \phi \phi / \Lambda_3^3 $ cubic mixing between $\phi$ and the metric and generates a new $\phi^4$ contact vertex responsible for the analytic $2\to 2$ amplitude. 

\newpage 

\begin{widetext}
\FloatBarrier
\begin{figure}[h!t]
\centering
		\begin{tikzpicture}[baseline=-0.6cm]
			\begin{feynman}
				\vertex (a1);
				\vertex [below=0.5cm of a1] (b1);
				\vertex [below=0.5cm of b1] (c1);
				\vertex [left=0.5cm of a1] (a2);
				\vertex [below=0.5cm of a2] (b2);
				\vertex [above left=0.4cm of b2] (b2lt);
				\vertex [below left=0.4cm of b2] (b2lb);
				\vertex [above right=0.4cm of b2] (b2rt);
				\vertex [below right=0.4cm of b2] (b2rb);
				\vertex [below=0.5cm of b2] (c2);
				\vertex [left=0.5cm of a2] (a3);
				\vertex [below=0.5cm of a3] (b3);
				\vertex [below=0.5cm of b3] (c3);
				\vertex [left=0.5cm of a3] (a4);
				\vertex [below=0.5cm of a4] (b4);
				\vertex [below=0.5cm of b4] (c4);
 				\node [below=0.1cm of a2, blob];
				
				\diagram*{
				(a3) -- [scalar] (b2lt),
				(c3) -- [scalar] (b2lb),
				(a1) -- [scalar] (b2rt),
				(c1) -- [scalar] (b2rb),
				};				
			\end{feynman}		
		\end{tikzpicture}
		\qquad = \quad
		\begin{tikzpicture}[baseline=-0.6cm]
			\begin{feynman}
				\vertex (a1);
				\vertex [below=0.5cm of a1] (b1);
				\vertex [below=0.5cm of b1] (c1);
				\vertex [left=0.5cm of a1] (a2);
				\vertex [below=0.5cm of a2] (b2);
				\vertex [below=0.5cm of b2] (c2);
				\vertex [left=0.5cm of a2] (a3);
				\vertex [below=0.5cm of a3] (b3);
				\vertex [below=0.5cm of b3] (c3);
				\vertex [left=0.5cm of a3] (a4);
				\vertex [below=0.5cm of a4] (b4);
				\vertex [below=0.5cm of b4] (c4);
				\node at (-0.5,  -1.5) {$ \frac{ \G_{4,XX} }{\Lambda_3^{6}} $};
				
				\diagram*{
				(a3) -- [scalar] (b2),
				(c3) -- [scalar] (b2),
				(a1) -- [scalar] (b2),
				(c1) -- [scalar] (b2),
				};				
			\end{feynman}		
		\end{tikzpicture}
		\qquad + \qquad
 		\begin{tikzpicture}[baseline=-0.6cm]
			\begin{feynman}
				\vertex (a1);
				\vertex [below=0.5cm of a1] (b1);
				\vertex [below=0.5cm of b1] (c1);
				\vertex [left=0.5cm of a1] (a2);
				\vertex [below=0.5cm of a2] (b2);
				\vertex [below=0.5cm of b2] (c2);
				\vertex [left=0.8cm of a2] (a3);
				\vertex [below=0.5cm of a3] (b3);
				\vertex [below=0.5cm of b3] (c3);
				\vertex [left=0.5cm of a3] (a4);
				\vertex [below=0.5cm of a4] (b4);
				\vertex [below=0.5cm of b4] (c4);
 				\node at (-0.8,  -1.5) {$\frac{ \G_{3, X}^2 }{\Lambda_3^{6}}$};
				
				\diagram*{
				(a4) -- [scalar] (b3),
				(c4) -- [scalar] (b3),
				(b3) -- [scalar] (b2),
				(b2) -- [scalar] (a1),
				(b2) -- [scalar] (c1),
				};				
			\end{feynman}		
		\end{tikzpicture}
		\qquad + \qquad
 		\begin{tikzpicture}[baseline=-0.6cm]
			\begin{feynman}
				\vertex (a1);
				\vertex [below=0.5cm of a1] (b1);
				\vertex [below=0.5cm of b1] (c1);
				\vertex [left=0.5cm of a1] (a2);
				\vertex [below=0.5cm of a2] (b2);
				\vertex [left=0.15cm of b2] (b2l);
				\vertex [above right=0.2cm of b2] (b2rt);
				\vertex [below right=0.2cm of b2] (b2rb);
				\vertex [below=0.5cm of b2] (c2);
				\vertex [left=0.8cm of a2] (a3);
				\vertex [below=0.5cm of a3] (b3);
				\vertex [right=0.15cm of b3] (b3r);
				\vertex [above left=0.2cm of b3] (b3lt);
				\vertex [below left=0.2cm of b3] (b3lb);
				\vertex [below=0.5cm of b3] (c3);
				\vertex [left=0.5cm of a3] (a4);
				\vertex [below=0.5cm of a4] (b4);
				\vertex [below=0.5cm of b4] (c4);
 				\node at (-0.8,  -1.5) {$\frac{ \G_{4,X}^2 }{\Lambda_3^{6}}$};
				
				\diagram*{
				(a4) -- [scalar] (b3),
				(c4) -- [scalar] (b3),
				(b3) -- [photon] (b2),
				(b2) -- [scalar] (a1),
				(b2) -- [scalar] (c1),
				};				
			\end{feynman}		
		\end{tikzpicture}
		\qquad + \qquad ...  \\
		\qquad\qquad + \qquad
		\begin{tikzpicture}[baseline=-0.6cm]
			\begin{feynman}
				\vertex (a1);
				\vertex [below=0.5cm of a1] (b1);
				\vertex [below=0.5cm of b1] (c1);
				\vertex [left=0.5cm of a1] (a2);
				\vertex [below=0.5cm of a2] (b2);
				\node   [below=0.425cm of a2, dot] (d);
				\vertex [below=0.5cm of b2] (c2);
				\vertex [left=0.5cm of a2] (a3);
				\vertex [below=0.5cm of a3] (b3);
				\vertex [below=0.5cm of b3] (c3);
				\vertex [left=0.5cm of a3] (a4);
				\vertex [below=0.5cm of a4] (b4);
				\vertex [below=0.5cm of b4] (c4);
				\node at (-0.5,  -1.5) {$\frac{ \G_{2, XX} }{M_P \Lambda_3^{3}}$};
				
				\diagram*{
				(a3) -- [scalar] (b2),
				(c3) -- [scalar] (b2),
				(a1) -- [scalar] (b2),
				(c1) -- [scalar] (b2),
				};				
			\end{feynman}		
		\end{tikzpicture}
		\qquad + \qquad
 		\begin{tikzpicture}[baseline=-0.6cm]
			\begin{feynman}
				\vertex (a1);
				\vertex [below=0.5cm of a1] (b1);
				\vertex [below=0.5cm of b1] (c1);
				\vertex [left=0.5cm of a1] (a2);
				\vertex [below=0.5cm of a2] (b2);
				\node   [below=0.425cm of a2, dot] (d);
				\vertex [left=0.15cm of b2] (b2l);
				\vertex [above right=0.2cm of b2] (b2rt);
				\vertex [below right=0.2cm of b2] (b2rb);
				\vertex [below=0.5cm of b2] (c2);
				\vertex [left=0.8cm of a2] (a3);
				\vertex [below=0.5cm of a3] (b3);
				\vertex [right=0.15cm of b3] (b3r);
				\vertex [above left=0.2cm of b3] (b3lt);
				\vertex [below left=0.2cm of b3] (b3lb);
				\vertex [below=0.5cm of b3] (c3);
				\vertex [left=0.5cm of a3] (a4);
				\vertex [below=0.5cm of a4] (b4);
				\vertex [below=0.5cm of b4] (c4);
 				\node at (-0.8,  -1.5) {$\frac{ 0 }{ M_P \Lambda_3^{3}}$};
				
				\diagram*{
				(a4) -- [scalar] (b3),
				(c4) -- [scalar] (b3),
				(b3) -- [scalar] (b2),
				(b2) -- [scalar] (a1),
				(b2) -- [scalar] (c1),
				};				
			\end{feynman}		
		\end{tikzpicture}
		\qquad + \qquad
 		\begin{tikzpicture}[baseline=-0.6cm]
			\begin{feynman}
				\vertex (a1);
				\vertex [below=0.5cm of a1] (b1);
				\vertex [below=0.5cm of b1] (c1);
				\vertex [left=0.5cm of a1] (a2);
				\vertex [below=0.5cm of a2] (b2);
				\node   [below=0.425cm of a2, dot] (d);
				\vertex [left=0.15cm of b2] (b2l);
				\vertex [above right=0.2cm of b2] (b2rt);
				\vertex [below right=0.2cm of b2] (b2rb);
				\vertex [below=0.5cm of b2] (c2);
				\vertex [left=0.8cm of a2] (a3);
				\vertex [below=0.5cm of a3] (b3);
				\vertex [right=0.15cm of b3] (b3r);
				\vertex [above left=0.2cm of b3] (b3lt);
				\vertex [below left=0.2cm of b3] (b3lb);
				\vertex [below=0.5cm of b3] (c3);
				\vertex [left=0.5cm of a3] (a4);
				\vertex [below=0.5cm of a4] (b4);
				\vertex [below=0.5cm of b4] (c4);
 				\node at (-0.8,  -1.5) {$\frac{ \G_{4,X} \G_{2,X} }{M_P \Lambda_3^{3}}$};
				
				\diagram*{
				(a4) -- [scalar] (b3),
				(c4) -- [scalar] (b3),
				(b3) -- [photon] (b2),
				(b2) -- [scalar] (a1),
				(b2) -- [scalar] (c1),
				};				
			\end{feynman}		
		\end{tikzpicture}
		\qquad + \qquad ... 
 \caption{ $\phi \phi \to \phi \phi$ scattering amplitude for Horndeski using the above vertices (below each diagram we include only the shift-symmetric parts of the $G_n$ for brevity), where $+...$ includes all permutations of external legs. The top line is the leading order result, while the bottom line is subleading in $M_P$ (and violates the Galileon symmetry, leading to a non-zero $c_{ss}^{\phi\phi}$).
\label{fig:feyn1} 
 }
\end{figure}
\begin{figure}[h!t]
\centering
		\begin{tikzpicture}[baseline=-0.6cm]
			\begin{feynman}
				\vertex (a1);
				\vertex [below=0.5cm of a1] (b1);
				\vertex [below=0.5cm of b1] (c1);
				\vertex [left=0.5cm of a1] (a2);
				\vertex [below=0.5cm of a2] (b2);
				\vertex [above left=0.4cm of b2] (b2lt);
				\vertex [below left=0.4cm of b2] (b2lb);
				\vertex [above right=0.4cm of b2] (b2rt);
				\vertex [below right=0.4cm of b2] (b2rb);
				\vertex [below=0.5cm of b2] (c2);
				\vertex [left=0.5cm of a2] (a3);
				\vertex [below=0.5cm of a3] (b3);
				\vertex [below=0.5cm of b3] (c3);
				\vertex [left=0.5cm of a3] (a4);
				\vertex [below=0.5cm of a4] (b4);
				\vertex [below=0.5cm of b4] (c4);
 				\node [below=0.1cm of a2, blob];
				
				\diagram*{
				(a3) -- [scalar] (b2lt),
				(c3) -- [photon] (b2lb),
				(a1) -- [scalar] (b2rt),
				(c1) -- [photon] (b2rb),
				};				
			\end{feynman}		
		\end{tikzpicture}
		\qquad = \quad
 		\begin{tikzpicture}[baseline=-0.6cm]
			\begin{feynman}
				\vertex (a1);
				\vertex [below=0.5cm of a1] (b1);
				\vertex [below=0.5cm of b1] (c1);
				\vertex [left=0.5cm of a1] (a2);
				\vertex [below=0.5cm of a2] (b2);
				\vertex [left=0.15cm of b2] (b2l);
				\vertex [above right=0.2cm of b2] (b2rt);
				\vertex [below right=0.2cm of b2] (b2rb);
				\vertex [below=0.5cm of b2] (c2);
				\vertex [left=0.8cm of a2] (a3);
				\vertex [below=0.5cm of a3] (b3);
				\vertex [right=0.15cm of b3] (b3r);
				\vertex [above left=0.2cm of b3] (b3lt);
				\vertex [below left=0.2cm of b3] (b3lb);
				\vertex [below=0.5cm of b3] (c3);
				\vertex [left=0.5cm of a3] (a4);
				\vertex [below=0.5cm of a4] (b4);
				\vertex [below=0.5cm of b4] (c4);
 				\node at (-0.8,  -1.5) {${\color{blue} \frac{ \G_{4,X}^2 }{\Lambda_3^{6}} }$};
				
				\diagram*{
				(a4) -- [scalar] (b3),
				(c4) -- [photon] (b3),
				(b3) -- [scalar] (b2),
				(b2) -- [scalar] (a1),
				(b2) -- [photon] (c1),
				};				
			\end{feynman}		
		\end{tikzpicture}
		\qquad + \qquad ...  \hspace{10cm} \hskip 10pt \\
		\qquad\qquad + \qquad
		\begin{tikzpicture}[baseline=-0.6cm]
			\begin{feynman}
				\vertex (a1);
				\vertex [below=0.5cm of a1] (b1);
				\vertex [below=0.5cm of b1] (c1);
				\vertex [left=0.5cm of a1] (a2);
				\vertex [below=0.5cm of a2] (b2);
				\node   [below=0.425cm of a2, dot] (d);
				\vertex [below=0.5cm of b2] (c2);
				\vertex [left=0.5cm of a2] (a3);
				\vertex [below=0.5cm of a3] (b3);
				\vertex [below=0.5cm of b3] (c3);
				\vertex [left=0.5cm of a3] (a4);
				\vertex [below=0.5cm of a4] (b4);
				\vertex [below=0.5cm of b4] (c4);
				\node at (-0.5,  -1.5) {$\frac{ \G_{4,X} }{M_P \Lambda_3^{3}}$};
				
				\diagram*{
				(a3) -- [scalar] (b2),
				(c3) -- [photon] (b2),
				(a1) -- [scalar] (b2),
				(c1) -- [photon] (b2),
				};				
			\end{feynman}		
		\end{tikzpicture}
		\qquad + \qquad
 		\begin{tikzpicture}[baseline=-0.6cm]
			\begin{feynman}
				\vertex (a1);
				\vertex [below=0.5cm of a1] (b1);
				\vertex [below=0.5cm of b1] (c1);
				\vertex [left=0.5cm of a1] (a2);
				\vertex [below=0.5cm of a2] (b2);
				\node   [below=0.425cm of a2, dot] (d);
				\vertex [left=0.15cm of b2] (b2l);
				\vertex [above right=0.2cm of b2] (b2rt);
				\vertex [below right=0.2cm of b2] (b2rb);
				\vertex [below=0.5cm of b2] (c2);
				\vertex [left=0.8cm of a2] (a3);
				\vertex [below=0.5cm of a3] (b3);
				\vertex [right=0.15cm of b3] (b3r);
				\vertex [above left=0.2cm of b3] (b3lt);
				\vertex [below left=0.2cm of b3] (b3lb);
				\vertex [below=0.5cm of b3] (c3);
				\vertex [left=0.5cm of a3] (a4);
				\vertex [below=0.5cm of a4] (b4);
				\vertex [below=0.5cm of b4] (c4);
 				\node at (-0.8,  -1.5) {${\color{blue} \frac{ \G_{4,X} \G_{2,X} }{ M_P \Lambda_3^{3}} }$};
				
				\diagram*{
				(a4) -- [scalar] (b3),
				(c4) -- [photon] (b3),
				(b3) -- [scalar] (b2),
				(b2) -- [scalar] (a1),
				(b2) -- [photon] (c1),
				};				
			\end{feynman}		
		\end{tikzpicture}
		\qquad + \qquad
 		\begin{tikzpicture}[baseline=-0.6cm]
			\begin{feynman}
				\vertex (a1);
				\vertex [below=0.5cm of b1] (c1);
				\vertex [left=0.5cm of a1] (a2);
				\vertex [below=0.5cm of a2] (b2);
				\vertex [below=0.5cm of b2] (c2);
				\vertex [left=0.5cm of a2] (a3);
				\vertex [below=0.5cm of a3] (b3);
				\vertex [below=0.5cm of b3] (c3);
				\vertex [left=0.5cm of a3] (a4);
				\vertex [below=0.5cm of a4] (b4);
				\vertex [below=0.5cm of b4] (c4);
				\vertex [below=0.25cm of a2] (a2l);
				\vertex [above=0.25cm of c2] (c2u);
 				\node at (-0.5,  -1.5) {${\color{blue} \frac{ 0 }{M_P \Lambda_3^{3}} }$};
 				\node   [above=0.15cm of c2, dot] (d);
				
				\diagram*{
				(a3) -- [scalar] (a2l) -- [scalar] (a1),
				(c3) -- [photon] (c2u) -- [photon] (c1),
				(c2u) -- [scalar] (a2l),
				};				
			\end{feynman}		
		\end{tikzpicture}
		\qquad + \qquad
 		\begin{tikzpicture}[baseline=-0.6cm]
			\begin{feynman}
				\vertex (a1);
				\vertex [below=0.5cm of b1] (c1);
				\vertex [left=0.5cm of a1] (a2);
				\vertex [below=0.5cm of a2] (b2);
				\vertex [below=0.5cm of b2] (c2);
				\vertex [left=0.5cm of a2] (a3);
				\vertex [below=0.5cm of a3] (b3);
				\vertex [below=0.5cm of b3] (c3);
				\vertex [left=0.5cm of a3] (a4);
				\vertex [below=0.5cm of a4] (b4);
				\vertex [below=0.5cm of b4] (c4);
				\vertex [below=0.25cm of a2] (a2l);
				\vertex [above=0.25cm of c2] (c2u);
 				\node at (-0.5,  -1.5) {$\frac{ \G_{4,X} \G_{4} }{M_P \Lambda_3^{3}}$};
 				\node   [above=0.15cm of c2, dot] (d);
				
				\diagram*{
				(a3) -- [scalar] (a2l) -- [scalar] (a1),
				(c3) -- [photon] (c2u) -- [photon] (c1),
				(c2u) -- [photon] (a2l),
				};				
			\end{feynman}		
		\end{tikzpicture}
		\qquad + \qquad ... 
 \caption{ $\phi h \to \phi h$ scattering amplitude for Horndeski using the above vertices (below each diagram we include only the shift-symmetric part of $G_n$ for brevity), where $+...$ includes all permutations of external legs. The diagrams in blue vanish when the external gravitons are taken on-shell, and the remaining two terms in black exactly cancel.
\label{fig:feyn2} 
 }
\end{figure}

\FloatBarrier
\end{widetext}

%%%%%%%%
%\bibliographystyle{utcaps}
\bibliographystyle{apsrev4-1}
\bibliography{pos_MCMC_v3}
%%%%%%%%

\end{document}